\documentclass[11pt]{article}
\usepackage{epsfig} 
\newcommand{\half}{\mbox{\small{$\frac{1}{2}$}}}

\newcommand{\MSbar}{\overline{\mbox{MS}}} 

\begin{document}
\title{Two loop $\MSbar$ renormalization of the Curci-Ferrari model} 
\author{J.A. Gracey, \\ Theoretical Physics Division, \\ Department 
of Mathematical Sciences, \\ University of Liverpool, \\ Peach Street, \\ 
Liverpool, \\ L69 7ZF, \\ United Kingdom.} 
\date{November, 2001.} 
\maketitle 
\vspace{5cm} 
\noindent 
{\bf Abstract.} We renormalize the Curci-Ferrari model at two loops in the 
$\MSbar$ scheme in an arbitrary covariant gauge. 

\vspace{-16cm} 
\hspace{13.5cm} 
{\bf LTH 530} 

\newpage 
The Curci-Ferrari model is an extension of Yang-Mills theory where the gluon
field, $A^a_\mu$, is massive, \cite{1,2}. However, as is well known in such a
massive non-abelian gauge theory, where there is no spontaneous symmetry 
breaking, the theory is not fully consistent. For instance, it does not obey
unitarity, \cite{2,3,4}, though it has been shown to be multiplicatively
renormalizable to all orders in perturbation theory, \cite{1,4,5,6}. Further, 
as the mass for the gluon is included in the Lagrangian in the simplest 
possible manner, there is the possibility that gauge invariance is also broken.
However, as was shown in the original work of Curci and Ferrari, \cite{1,2}, if
one also includes a particular form of mass term for the ghost fields when the 
gauge is fixed covariantly in a non-linear way, the mass term in the Lagrangian
in fact preserves BRST symmetry, though without restoring unitarity which 
remains broken. Despite the lack of unitarity the Curci-Ferrari model has been 
of interest recently for a variety of reasons. First, there have been several
investigations into how a mass gap for the gluon can arise, \cite{7,8,9}. For 
example, in \cite{8,9} a gluon and ghost mass term akin to that which occurs in
the Curci-Ferrari model has been argued to be important in generating a mass 
gap in ordinary Yang-Mills theories. For instance, a non-zero vacuum 
expectation value for $\langle \half (A_\mu^a)^2 ~+~ \alpha \bar{c}^a c^a
\rangle$ emerges under certain assumptions where $\alpha$ is the covariant
gauge parameter and $c^a$ and $\bar{c}^a$ are the respective ghost and 
anti-ghost fields. Such an operator represents the simplest {\em local}
dimension two operator which can be constructed in non-abelian gauge field 
theories. Also, a similar approach has been considered in \cite{7} where the 
effective potential of the Landau gauge condensate is constructed. There it was 
shown that at two loops the non-perturbative vacuum favoured a non-zero value 
for the vacuum expectation value. Other studies include \cite{10}. Second, the 
model itself has provided a useful laboratory for studying gauge theories with 
a non-linear gauge fixing, \cite{5,11}. Third, massive Yang-Mills models have 
been used to model the phenomenology of the strong interactions. For example, 
such a model was used in \cite{12} to try and understand diffractive 
scattering. Finally, the main motivation for our interest in the Curci-Ferrari 
model stems from the role of the gluon mass as a {\em natural} infrared 
regulator for Feynman integrals which arise in loop calculations, 
\cite{13,6,14}. 

In the renormalization of QCD at high loop orders the standard method of
extracting the ultraviolet divergence structure of a Feynman integral with
respect to the regularization, such as dimensional regularization, is to 
perform the calculation with massless propagators. However, for two and higher 
loop integrals one has to be careful that spurious infrared infinities are not
generated. These can occur when one reduces the class of Feynman integrals to 
vacuum integrals by expanding in powers of the external momenta after one or
more external momenta have been set to zero first. To circumvent this potential 
problem the technique of infrared rearrangement, \cite{15,16}, is applied where
a temporary mass is added to the appropriate propagators in an infrared 
divergent integral. Whilst this has been a hugely successful technique it 
suffers from the drawback that it is performed by hand and hence limited when 
considering the large number of diagrams which will arise when the loop order 
increases. Moreover, it was not clear until recently how one could develop the 
method for application in an automatic multiloop computer algebra programme. 
One recent approach to address this problem is that of \cite{17,18}. There each
propagator is systematically given an infrared cutoff mass so that the 
resulting vacuum integrals are automatically infrared finite. Provided all the 
vacuum diagrams are computable at that loop order then one can automatize the 
process. Indeed \cite{17,18} provides a full two loop calculation of various 
anomalous dimensions. Given that the Curci-Ferrari model {\em naturally} 
incorporates a mass akin to that introduced for infrared rearrangement and 
{\em remains} renormalizable it seems appropriate to consider and develop that 
model as a useful and alternative tool to compute the ultraviolet structure of 
Yang-Mills theories. This is important since the efficient programmes such as 
{\sc Mincer}, \cite{19}, compute the ultraviolet structure of only massless 
two-point functions at three loops, but for $n$-point functions with 
$n$~$>$~$2$ one cannot naively nullify all bar two of the external momenta, 
similar to infrared rearrangement, and apply {\sc Mincer}. This is because this
procedure will inevitably give rise to spurious infrared infinities which, in 
dimensional regularization, cannot be distinguished from the desired 
ultraviolet divergences. Hence, it would seem more appropriate to us to apply 
the Curci-Ferrari model in these circumstances. However, it turns out that 
currently the model has only been renormalized to one loop in \cite{4}. 
Moreover, an earlier calculation, \cite{14}, appears to incorrectly determine 
the renormalization constants which must satisfy the appropriate Slavnov-Taylor
identities. Therefore, the purpose of this letter is to renormalize the 
Curci-Ferrari model at two loops in the $\MSbar$ scheme using an arbitrary 
(non-linear) covariant gauge fixing term and dimensional regularization. This 
is necessary given that the renormalization constants of the fields, coupling, 
gauge parameter and masses are required for the subsequent renormalization of 
any Green's function. As a consequence of our calculations we will verify the 
result of \cite{4} that five renormalization constants are necessary to render 
the model finite as opposed to the three suggested in \cite{3}. 

The Lagrangian of the Curci-Ferrari model is, \cite{1},  
\begin{eqnarray} 
L &=& -~ \frac{1}{4} G_{\mu\nu}^a G^{a \, \mu\nu} ~-~ \frac{1}{2\alpha} 
(\partial^\mu A^a_\mu)^2 ~-~ \frac{m^2}{2} A_\mu^a A^{a \, \mu} ~+~ 
\partial_\mu \bar{c}^a \partial^\mu c^a \nonumber \\ 
&& -~ \alpha m^2 \bar{c}^a c^a ~-~ \frac{g}{2} f^{abc} A^a_\mu \, \bar{c}^b \, 
{\stackrel \leftrightarrow {\partial^\mu} } \, c^c ~+~ \frac{\alpha g^2}{8} 
f^{eab} f^{ecd} \bar{c}^a c^b \bar{c}^c c^d 
\label{lag}
\end{eqnarray}  
where $1$ $\leq$ $a$ $\leq$ $N_{\! A}$ with $N_{\! A}$ the dimension of the
adjoint representation, $D_\mu$ $=$ $\partial_\mu$ $+$ $ig T^a A^a_\mu$, 
$G^a_{\mu\nu}$~$=$~$\partial_\mu A^a_\nu$~$-$~$\partial_\nu A^a_\mu$~$-$~$g 
f^{abc} A^b_\mu A^c_\nu$, $g$ is the coupling constant and $m$ is the gluon 
mass. The ghost mass is related to the gluon mass by the covariant gauge fixing
parameter $\alpha$. The structure constants of the Lie group are $f^{abc}$ and 
$\bar{c}^a \, {\stackrel \leftrightarrow {\partial_\mu} } \, c^b$ $=$ 
$\bar{c}^a \partial_\mu c^b$ $-$ $(\partial_\mu \bar{c}^a) c^b$. Following the 
usual procedures the gluon and ghost propagators are respectively  
\begin{equation}
-~ \delta^{ab} \left[ \frac{\eta^{\mu\nu}}{(k^2+m^2)} ~-~ \frac{(1-\alpha)k^\mu
k^\nu}{(k^2+m^2)(k^2+\alpha m^2)} \right] ~~~~ \mbox{and} ~~~~ 
\frac{\delta^{ab}}{(k^2+\alpha m^2)} ~.  
\label{propdefn} 
\end{equation} 
The unphysical pole at $\alpha m^2$ in the ghost propagator had led to the hope
that it would counteract the same pole in the gluon propagator to establish a 
unitary theory. However, it has been shown, \cite{1,2,3,4}, that this is not
the case. To renormalize (\ref{lag}) the quantities of the bare Lagrangian,
denoted by the subscript $\mbox{o}$, are replaced by the renormalized ones, 
through,  
\begin{eqnarray} 
A^{a \, \mu}_{\mbox{\footnotesize{o}}} &=& \sqrt{Z_A} \, A^{a \, \mu} ~~,~~ 
c^a_{\mbox{\footnotesize{o}}} ~=~ \sqrt{Z_c} \, c^a ~~,~~ 
\bar{c}^a_{\mbox{\footnotesize{o}}} ~=~ \sqrt{Z_c} \, \bar{c}^a \nonumber \\ 
g_{\mbox{\footnotesize{o}}} &=& Z_g \, g ~~,~~ m_{\mbox{\footnotesize{o}}} ~=~ 
Z_m \, m ~~,~~ \alpha_{\mbox{\footnotesize{o}}} ~=~ Z^{-1}_\alpha Z_A \, 
\alpha ~. 
\label{Zdefns}
\end{eqnarray} 
We have assumed initially that there are five independent renormalization
constants. In ordinary Yang-Mills one has three renormalization constants since
clearly there is no need for $Z_m$ and the Slavnov-Taylor identities ensure 
that $Z_\alpha$ $=$ $1$ in our notation to all orders to preserve gauge 
symmetry. Here we will allow for an independent $Z_\alpha$ as in \cite{4}. To 
find each of the renormalization constants to two loops the gluon and ghost 
two-point functions are computed first which determine $Z_A$, $Z_\alpha$, $Z_c$
and $Z_m$. It is worth pointing out that since the gluon two-point function 
gives $Z_A$, $Z_\alpha$ and $Z_m$ the structure of the ghost mass 
renormalization is determined prior to calculating it. This in fact provides a 
strong check on our results. In addition we have computed the ghost-gluon and 
triple gluon vertex renormalizations to verify that the correct scheme 
independent two loop $\beta$-function emerges in both the gluon and ghost 
sectors. The renormalization group functions are defined by, (see, for example,
\cite{20}), 
\begin{eqnarray} 
\gamma_A(a) &=& \mu \frac{\partial \ln Z_A}{\partial \mu} ~~,~~  
\gamma_c(a) ~=~ \mu \frac{\partial \ln Z_c}{\partial \mu} ~~,~~  
\beta(a) ~=~ \mu \frac{\partial a}{\partial \mu} \nonumber \\  
\gamma_m(a) &=& \frac{\partial \ln m}{\partial \ln \mu} ~~,~~ 
\gamma_\alpha(a) ~=~ \frac{\partial \ln \alpha}{\partial \ln \mu} 
\end{eqnarray} 
where $a$ $=$ $g^2/(16\pi^2)$. From (\ref{Zdefns}) these imply 
\begin{eqnarray} 
\gamma_A(a) &=& \beta(a) \frac{\partial \ln Z_A}{\partial a} ~+~
\alpha \gamma_\alpha(a) \frac{\partial \ln Z_A}{\partial \alpha} \nonumber \\
\gamma_\alpha(a) &=& \left[ \beta(a) \frac{\partial \ln Z_\alpha}{\partial a} 
{}~-~ \gamma_A(a) \right] \left[ 1 ~-~ \alpha \frac{\partial 
\ln Z_\alpha}{\partial \alpha} \right]^{-1} \nonumber \\ 
\gamma_m(a) &=& -~ \mu \frac{\partial \ln Z_m}{\partial \mu} ~. 
\label{rgedefn} 
\end{eqnarray} 
Ordinarily in massless Yang-Mills theory one has the simple relation 
$\gamma_\alpha(a)$ $=$ $-$ $\gamma_A(a)$. When the condition $Z_\alpha$ $=$ $1$
is not satisfied the more general relation, (\ref{rgedefn}), emerges.

To compute the ultraviolet structure of the two and three point functions we
have followed the strategy of \cite{17,18} of expanding out the Feynman 
integrals to the terms involving two or three external momenta respectively. 
Given that the Curci-Ferrari model is renormalizable, \cite{1,4,5,6}, it 
follows that terms involving more external momenta will not contribute to the 
ultraviolet divergences. Hence one is left with massive vacuum two loop 
Feynman integrals to compute where, unlike the calculation of \cite{17,18}, the
masses in the propagators are not all the same. Such integrals have been widely
studied before and we quote the result of \cite{21}, for instance, for the 
basic two loop topology. In general we have 
\begin{eqnarray} 
\int_{kl} \frac{1}{(k^2+m_1^2)(l^2+m_2^2)[(k-l)^2+m_3^2]} &=& -~ 
(m_1^2+m_2^2+m_3^2) \left( \frac{1}{2\epsilon^2} ~+~ \frac{3}{2\epsilon} ~+~  
\frac{1}{\epsilon} \ln(4\pi e^{-\gamma}) \right) \nonumber \\
&& +~ \left( m_1^2 \ln m_1^2 ~+~ m_2^2 \ln m_2^2 ~+~ m_3^2 \ln m_3^2 \right) 
\frac{1}{\epsilon} ~+~ O(1) \nonumber \\  
\end{eqnarray}
for arbitrary masses $m_i$ where $\int_k$ $=$ $\int d^d k/(2\pi)^d$, $d$ $=$ 
$4$ $-$ $2\epsilon$ with $\epsilon$ the regularizing parameter and $\gamma$ is
the Euler-Mascheroni constant. The integrals involving different powers of a 
propagator are given by differentiating with respect to the appropriate masses.
It is important to realise the role of the logarithmic terms. For instance, 
specifying various values for $m^2_i$ we have  
\begin{eqnarray} 
\int_{kl} \frac{1}{(k^2+m^2)(l^2+m^2)[(k-l)^2+\alpha m^2]} &=& -~ (\alpha+2) 
m^2 \left( \frac{1}{2\epsilon^2} ~+~ \frac{3}{2\epsilon} ~+~ \frac{1}{\epsilon}
\ln(4\pi e^{-\gamma})\right) \nonumber \\
&& +~ \left( (\alpha+2) \ln m^2 ~+~ \alpha \ln \alpha \right) 
\frac{m^2}{\epsilon} ~+~ O(1) 
\end{eqnarray}
and
\begin{eqnarray} 
\int_{kl} \frac{1}{(k^2+m^2)(l^2+ \alpha m^2)[(k-l)^2+\alpha m^2]} &=& -~ 
(2\alpha+1) m^2 \left( \frac{1}{2\epsilon^2} ~+~ \frac{3}{2\epsilon} ~+~ 
\frac{1}{\epsilon} \ln(4\pi e^{-\gamma}) \right) 
\nonumber \\
&& +~ \left( (2\alpha+1) \ln m^2 ~+~ 2 \alpha \ln \alpha \right) 
\frac{m^2}{\epsilon} ~+~ O(1) ~. \nonumber \\ 
\label{basint} 
\end{eqnarray}
Therefore, the final renormalization constants could potentially contain 
$\ln \alpha$ terms. Moreover, in writing two and three point functions in the 
basic form (\ref{basint}), one uses partial fractions so that  
\begin{equation} 
\frac{1}{(k^2+m^2)(k^2+\alpha m^2)} ~=~ \frac{1}{(1-\alpha)m^2} 
\left[ \frac{1}{(k^2+\alpha m^2)} ~-~ \frac{1}{(k^2+m^2)} \right] 
\end{equation} 
which introduces, in addition, powers of $1/(1-\alpha)$ which also could appear
in the renormalization constants. Given that $\ln\alpha$ and $1/(1-\alpha)^n$
for $n$ $\geq$ $1$ are singular at $\alpha$ $=$ $0$ and $\alpha$ $=$ $1$
respectively, it might be expected that at two loops there will be a problem in
the Landau and Feynman gauges respectively. However, since the original gluon
propagator, (\ref{propdefn}), can be rewritten as  
\begin{equation}
-~ \delta^{ab} \left[ \frac{\eta^{\mu\nu}}{(k^2+m^2)} ~-~ 
\frac{k^\mu k^\nu}{m^2} \left( \frac{1}{(k^2+\alpha m^2)} ~-~ 
\frac{1}{(k^2+m^2)} \right) \right] 
\end{equation} 
there ought to be no problems at $\alpha$ $=$ $1$ which provides an internal
consistency check on the computation. Moreover, the potential $\alpha$ $=$ $0$
singularity is avoided by the fact that the basic one loop integral has the 
following $\epsilon$-expansion 
\begin{equation} 
\int_k \frac{1}{(k^2+\alpha m^2)} ~=~ -~ \frac{\alpha m^2}{\epsilon} \left[
1 ~+~ [1 - \gamma - \ln (\alpha m^2) ] \epsilon ~+~ O(\epsilon^2) \right] ~.  
\end{equation} 
Hence, the counterterms which arise from the one loop diagrams ought to cancel 
out the $\ln(\alpha)/\epsilon$ poles arising in integrals of the type we have 
discussed. Indeed in this context we avoid the usual approach of subtractions 
by using the method of \cite{22}. There one computes the Green's functions in 
terms of the bare parameters and then rescales them at the end of the 
calculation in terms of the coupling constant expansion of the renormalization 
constants, (\ref{Zdefns}), where each term of the expansion has already been 
determined. This strategy is appropriate given that we have carried out the 
calculation automatically using a symbolic manipulation programme written in 
{\sc Form}, \cite{23}, in order to handle the tedious amount of algebra. The 
Feynman diagrams for such an approach were generated using {\sc Qgraf}, 
\cite{24}. 

Consequently, we find the following $\MSbar$ renormalization constants for the 
Curci-Ferrari model, (\ref{lag}),  
\begin{eqnarray} 
Z_A &=& 1 ~+~ C_A \left( \frac{13}{6} - \frac{\alpha}{2} \right) 
\frac{a}{\epsilon} \nonumber \\
&& +~ C_A^2 \left[ \left( \frac{3\alpha^2}{16} - \frac{17\alpha}{24} 
- \frac{13}{8} \right) \frac{1}{\epsilon^2} ~-~ \left( \frac{\alpha^2}{16} 
+ \frac{11\alpha}{16} - \frac{59}{16} \right) \frac{1}{\epsilon} 
\right] a^2 ~+~ O(a^3) \nonumber \\ 
Z_\alpha &=& 1 ~-~ C_A \left( \frac{\alpha}{4} \right) 
\frac{a}{\epsilon} \nonumber \\
&& +~ C_A^2 \left[ \left( \frac{\alpha^2}{16} + \frac{3\alpha}{16} 
\right) \frac{1}{\epsilon^2} ~-~ \left( \frac{\alpha^2}{32} 
+ \frac{5\alpha}{32} \right) \frac{1}{\epsilon} \right] a^2 ~+~ O(a^3) 
\nonumber \\ 
Z_c &=& 1 ~+~ C_A \left( \frac{3}{4} - \frac{\alpha}{4} \right) 
\frac{a}{\epsilon} \nonumber \\
&& +~ C_A^2 \left[ \left( \frac{\alpha^2}{16} - \frac{35}{32} \right) 
\frac{1}{\epsilon^2} ~-~ \left( \frac{\alpha^2}{32} - \frac{\alpha}{32} 
- \frac{95}{96} \right) \frac{1}{\epsilon} \right] a^2 ~+~ O(a^3) \nonumber \\ 
Z_m &=& 1 ~+~ C_A \left( \frac{\alpha}{8} - \frac{35}{24} \right) 
\frac{a}{\epsilon} \nonumber \\
&& +~ C_A^2 \left[ \left( - \frac{\alpha^2}{128} - \frac{53\alpha}{192} 
+ \frac{1435}{384} \right) \frac{1}{\epsilon^2} ~+~ 
\left( \frac{\alpha^2}{64} + \frac{11\alpha}{64} - \frac{449}{192} \right) 
\frac{1}{\epsilon} \right] a^2 ~+~ O(a^3) \nonumber \\ 
Z_g &=& 1 ~-~ \frac{11}{6} C_A \frac{a}{\epsilon} ~+~ C_A^2 \left[ 
\frac{121}{24} \frac{1}{\epsilon^2} ~-~ \frac{17}{6} \frac{1}{\epsilon} \right]
a^2 ~+~ O(a^3) 
\end{eqnarray}  
where $f^{acd} f^{bcd}$ $=$ $C_A \delta^{ab}$. As an additional check on our 
two loop results we note that the $O(1/\epsilon^2)$ pole terms of each 
renormalization constant agrees with the value predicted through the 
renormalization group from the one loop pole. Using (\ref{rgedefn}), these 
values for the renormalization constants lead to the renormalization group 
functions 
\begin{eqnarray} 
\gamma_A(a) &=& C_A ( 3\alpha - 13 ) \frac{a}{6} ~+~ C_A^2 \left(
\alpha^2 + 11\alpha - 59 \right) \frac{a^2}{8} ~+~ O(a^3) \nonumber \\  
\gamma_\alpha(a) &=& -~ C_A ( 3\alpha - 26 ) \frac{a}{12} ~-~ 
C_A^2 \left( \alpha^2 + 17\alpha - 118 \right) \frac{a^2}{16} ~+~ O(a^3) 
\nonumber \\ 
\gamma_c(a) &=& C_A ( \alpha - 3 ) \frac{a}{4} ~+~ C_A^2 \left(
3\alpha^2 - 3\alpha - 95 \right) \frac{a^2}{48} ~+~ O(a^3) \nonumber \\  
\gamma_m(a) &=& C_A ( 3\alpha - 35 ) \frac{a}{24} ~+~ C_A^2 \left(
3\alpha^2 + 33\alpha - 449 \right) \frac{a^2}{96} ~+~ O(a^3) \nonumber \\
\beta(a) &=& -~ \frac{11}{3} C_A a^2 ~-~ \frac{34}{3} C_A^2 a^3 ~+~ O(a^4) ~.  
\end{eqnarray} 
The result for the $\beta$-function agrees with the original two loop scheme
independent result of \cite{25,26}. As the gluon mass in (\ref{lag}) 
corresponds to the coupling of a gauge variant, and therefore unphysical, 
operator its mass anomalous dimension is gauge dependent. Further, the result 
for $\gamma_m(a)$ in the Landau gauge agrees with the calculation of \cite{7}
as an additional check on our computation. It is also worth noting that the 
explicit sum of the gluon wave function and gauge fixing parameter anomalous
dimensions gives  
\begin{equation} 
\gamma_A(a) ~+~ \gamma_\alpha(a) ~=~ \alpha \left[ C_A \frac{a}{4} ~+~ C_A^2 
\left( \alpha + 5 \right) \frac{a^2}{16} \right] ~+~ O(a^3) ~.  
\end{equation} 
As a final check on our results we note that in the Landau gauge 
$\gamma_\alpha(a)$ $=$ $-$ $\gamma_A(a)$ as in the massless Yang-Mills theory
and it is trivial to verify that the Landau gauge renormalization group 
functions coincide precisely with the two loop results of \cite{25,26,27,22}. 
In this instance the gluon propagator, 
(\ref{propdefn}), takes the simple transverse form  
\begin{equation} 
-~ \frac{\delta^{ab}}{(k^2+m^2)} \left[ \eta^{\mu\nu} ~-~ \frac{k^\mu k^\nu}
{k^2} \right] ~.  
\end{equation} 
Further, if we define the ghost mass as   
\begin{equation} 
m^2_c ~=~ \alpha m^2 
\end{equation} 
then in general, using $Z_{m_c}$ $=$ $Z_m Z^{\half}_A Z_\alpha^{-\half}$,  
\begin{equation} 
\gamma_{m_c}(a) ~=~ \gamma_m(a) ~-~ \frac{1}{2} \gamma_A(a) ~+~ 
\frac{1}{2} \left[ \beta(a) \frac{\partial \ln Z_\alpha}{\partial a} ~+~
\alpha \gamma_\alpha(a) \frac{\partial \ln Z_\alpha}{\partial \alpha} \right] 
\end{equation}  
giving to two loops 
\begin{equation} 
\gamma_{m_c}(a) ~=~ -~ \frac{3}{8} C_A a ~-~ C_A^2 \left( 18 \alpha + 95 
\right) \frac{a^2}{96} ~+~ O(a^3) ~.  
\end{equation} 

To conclude we have explicitly constructed all the basic renormalization group
functions for the Curci-Ferrari model, (\ref{lag}), at two loops in $\MSbar$. 
These will be important for using this theory to renormalize Green's functions 
in Yang-Mills theories where one cannot readily apply current automatic 
multiloop programmes to determine the ultraviolet structure of $n$-point 
functions with $n$ $>$ $3$ and where an infrared regularization is also 
necessary. Further, given that QCD is more appropriate for practical 
phenomenology it would be interesting to extend our computations to include 
massive quarks in addition to the massive gluons and ghosts. Although that 
theory would remain non-unitary, multiplicative renormalizability ought still 
to be preserved and hence the quark extended Curci-Ferrari model would provide 
a useful tool for renormalizing full QCD.  

\vspace{0.3cm} 
\noindent 
{\bf Acknowledgements.} The author thanks Dr D.B. Ali for useful discussions.

\end{document}